\documentclass[preprint]{emulateapj}

\newcommand\beq{\begin{equation}}
\newcommand\eeq{\end{equation}}
\newcommand\beqar{\begin{eqnarray}}
\newcommand\eeqar{\end{eqnarray}}

\begin{document}

\title{The Spectral Shape of the Gamma-ray Background from Blazars }

\author{Vasiliki Pavlidou\altaffilmark{1} \& 
Tonia M. Venters\altaffilmark{2,3} }

\altaffiltext{1}{
Kavli Institute of Cosmological Physics and Enrico Fermi Institute, 
The University of Chicago, Chicago, IL 60637}
\altaffiltext{2}{Department of Astronomy and Astrophysics, The University 
of Chicago, Chicago, IL 60637}
\altaffiltext{3}{Laboratoire d'AstroParticule et Cosmologie, Universit{\'e} de Paris 7 - Denis Diderot, Paris 13$^{\mbox{\`e}me}$, France}

\begin{abstract}

The spectral shape of the unresolved emission from different classes
of gamma-ray emitters can be used to disentangle the contributions
from these populations to the extragalactic gamma-ray background
(EGRB).    
We present a calculation of the unabsorbed {\em spectral shape} of the
unresolved blazar contribution to the EGRB starting from 
the spectral
index distribution (SID) 
of resolved EGRET blazars derived through a maximum-likelihood
analysis accounting for measurement errors. In addition, we explicitly
calculate the {\em uncertainty} in this theoretically predicted
spectral shape, which enters through the spectral index distribution
parameters.  We find that: (a) the
unresolved blazar emission spectrum is only mildly convex, and thus, even
if blazars are shown by GLAST to be a dominant contribution to the
ERGB at lower energies, they may be insufficient to explain the EGRB
at higher energies; (b) the theoretically predicted unresolved
spectral shape involves significant uncertainties due to the limited
constraints provided by EGRET data on the SID parameters, which are comparable to the
statistical uncertainties of the observed EGRET EGRB at high
energies; (c) the increased number statistics which will be provided by
GLAST will be sufficient to reduce this uncertainty by at least a
factor of three. 
\end{abstract}

\maketitle 

\section{Introduction}

The isotropic, and presumably, extragalactic gamma-ray background
emission (EGRB) detected by the Energetic Gamma-ray Experiment
Telescope (EGRET) aboard the {\it Compton Gamma-ray Observatory}
(Sreekumar et al.\ 1998) is one of the most important observational
constraints on known or theorized populations of faint, unresolved
gamma-ray emitters. With the imminent launch of the {\it Gamma-ray
  Large Area Space Telescope} (GLAST), which is expected to represent
an  unprecedented leap in observational capabilities in GeV energies,
the timing is especially opportune to consider the information content
of the diffuse background and methods for maximizing the scientific
return from its study.  

One of the primary challenges in using the EGRB to constrain
properties of extragalactic gamma-ray emitters and exotic physics is
disentangling the convolved contributions of guaranteed participating
populations. Estimates of the levels of the collective unresolved
emission even from established classes of extragalactic sources (such
as blazars and normal galaxies) involve significant uncertainties and
are at the order-of-magnitude level at best (e.g., Padovani et al.\
1993; Stecker \& Salamon 1996; Kazanas \& Perlman 1997; Mukherjee \&
Chiang 1999; M\"ucke \& Pohl 2000; Dermer 2006; Lichti et al.\ 1978;
Pavlidou \& Fields 2002). 

A very promising approach for the study of the EGRB and its components
is through the use of spectral shape information. Let us consider the
optimal case where the expected spectral shapes of the unresolved
emission from known classes of gamma-ray sources can be confidently
predicted. In this case, a series of conclusions can be drawn
regarding the potential contributions of these classes to the EGRB
even without detailed calculations of the {\em magnitudes} of their
collective emission. 
For example, in comparing the spectral shape of the spectrum due to a
particular class with that of the EGRB, one can identify whether this
class could, in principle, comprise most of the EGRB or require the
existence of contributions from other classes (Stecker \& Salamon
1996a,b; Strong et al. 2004; Pavlidou \& Fields 2002)\footnote{Note
  that the method of spectral comparison can only be used to {\em
    reject} a population from being the sole source of the EGRB;
  spectral consistency does not constitute in itself proof of the
  importance of a class of objects as an EGRB contributor, since the
  overall normalization of the emission may, in fact, be low depending on the gamma-ray luminosity function of the population.}.
 Potentially
identifiable spectral features could be predicted and searched for in
GLAST data (e.g., Pavlidou \& Fields 2002). Finally, spectral
information could be used to ultimately disentangle different
components and contributions (as in e.g. de Boer et al.\ 2004 for the
case of the diffuse emission from the Milky Way). An additional
attractive feature of such calculations is that the associated
uncertainties are largely independent of those entering the
calculations of the overall unresolved emission flux. 
 
As blazars are the most populated class of gamma-ray emitters, unresolved blazars are guaranteed to contribute significantly, if not dominantly, to the EGRB. Thus, it is especially important to understand the expected spectral shape of their collective unresolved emission and the uncertainties involved in its calculation. 
Individual blazars have been measured to have power-law spectra in the EGRET range, $F_E \propto E^{-\alpha}$, with spectral index $\alpha$ ranging approximately from 1.5 to 3. The unresolved emission from a collection of power-law emitters with variable spectral indices\footnote{Statistically significant spread in the observed and intrinsic spectral index of blazars has also been confirmed in other energy bands (see e.g.\ Shen et al.\ 2006 for the case of X-ray emission).} has, invariably, a convex spectral shape (Brecher \& Burbidge 1972; Stecker \& Salamon 1996; Pohl et al.\ 1997; Pavlidou et al.\ 2007).
The exact shape of the unresolved spectrum depends on the spectral index distribution (SID) of gamma-ray loud blazars. 
Recognizing that this is the case, Stecker \& Salamon (1996a) explicitly reconstructed a spectrum from the observed SID of blazars in the 2nd EGRET catalog (Thompson et al.\ 1995), deriving a spectrum that was indeed significantly convex. However, measurement errors in individual spectral indices smear the SID and exaggerate the curvature of the spectrum (Pohl et al.\ 1997). 

Recently, in Venters \& Pavlidou 2007 (hereafter VP07),  we have
applied a maximum-likelihood analysis to recover the intrinsic
spectral index distribution (ISID) of gamma-ray loud blazars from
EGRET observations. We found that (1) the maximum-likelihood ISID is
appreciably narrower than the observed SID, so the 
best-guess spectrum is likely to have only a mild curvature; 
(2) BL Lacs and flat
spectrum radio quasars (FSRQs) are likely to be spectrally distinct
populations with spectrally distinct contributions to the EGRB; (3)
there is no evidence for a systematic shift of spectral variability
with flaring implying that although variability may be important in
the level of the contribution from blazars, ignoring variability
effects in spectral shape studies is likely to be a good
approximation. 

Here, we use the ISIDs derived in VP07 to calculate the spectral shape
of the blazar contribution to the EGRB. We examine the sensitivity of
the shape to the exact values of the ISID parameters and report on the
range of possible shapes given our uncertainties in the determination
of these parameters. We also investigate how the spectral shapes of
the BL Lac and FSRQ contributions may differ.  Finally, we predict how
our understanding of the spectral shape of the unresolved blazar
emission will improve after GLAST observations become available.  

\section{Formalism}  

 If the differential photon flux spectrum of a single blazar is
$F_E(E) = F_{E,0} (E/E_0)^{-\alpha}$ (photons per unit area per unit energy per unit time), then the total flux of photons with
energies $>E_0$ is $F(>E_0) = F_{E,0}E_0 /(\alpha-1)$ (photons per
unit area per unit time). The
contribution of a single unresolved blazar of $F(>E_0) =F$ to the
EGRB is
\begin{equation}
I_{1} = (\alpha-1)\frac{F}{4\pi E_0}\left(\frac{E}{E_0}\right)^{-\alpha}
\end{equation} 
where $I$ has units of photons per unit area per unit energy per unit time per unit solid
angle, and the flux of one source is uniformly distributed over
$4\pi$. 

Now let us assume that the flux distribution of unresolved blazars
(number of objects per flux interval) can be described by a function
$g(F)$; and that the distribution of spectral indices (number of
objects per spectral index interval) can be described by a function 
$p(\alpha)$. Then the total contribution of unresolved blazars to the
EGRB is
\begin{equation}\label{complex}
I_{\rm EGRB}(E) = 
\int_{\alpha = -\infty}^{\infty}d\alpha
\int_{F=0}^{F_{\rm min}}\!\!\!\!\!dF \, g(F) I_{1} p(\alpha)\,. \nonumber \\
\end{equation}
In Eq.\ (\ref{complex}), $F_{\rm min}$ is the minimum flux of an object
that can be resolved by the telescope under consideration. 

We now make the assumptions that (a) blazar spectra can be adequately described by single power-laws in the observed energy range, as well as at energies which redshift down to the observed range; (b) the flux distribution is independent of spectral index; and (c) the spectral index does not evolve with time and does not depend on luminosity.
In this case, blazars in any single flux interval sample an identical SID, and produce the same fraction of photons in any two energy bins, thus resulting in a unique spectral shape.  This is intuitively reasonable. Had the (non-evolving) ISID been a $\delta$-function, all blazars would be power laws of the same slope, independently of the epoch of observation and the coadded spectrum would identically be a power law of the same slope. Since our ISID is a Gaussian of non-evolving spectral indices, the spectral shape is curved, but still independent of the luminosity function. The blazars in a specific unresolved flux interval contributing to the background will represent a mixture of luminosities and redshifts,  but since the blazar properties that determine the flux interval (redshift, luminosity) do not depend on the spectral index, the blazars within that flux interval will fairly sample the same ISID.  Additionally, redshifting down the spectra has no effect on the slope for single power laws as long as no absorption occurs. The same reasoning is applicable to all flux intervals, which therefore contribute portions of the background that have different amplitude but the same, unique spectral shape, dependent only on the parameters of the ISID.

The magnitude and spectral shape factors in Eq.\ (\ref{complex}) therefore decouple under our assumptions, and Eq.\ (\ref{complex}) can be rewritten 
as  
\begin{equation}\label{tria}
I_{\rm EGRB}(E) = I_0 \int_{\alpha = -\infty}^{\infty}\!\!\!\!d\alpha 
(\alpha-1)\left(
\frac{E}{E_0}\right)^{-\alpha}\!\!\!\!\!p(\alpha)\,, 
\end{equation}
where $I_0$ is a normalization constant depending on the flux
distribution of unresolved blazars. 
If $p(\alpha)$ is Gaussian, as assumed in VP07, then Eq.\ (\ref
{tria}) is analytically integrable (Pavlidou et al.\ 2007). 

It should be noted that if the above assumptions do not hold, then the spectral shape may not decouple from its magnitude as above.  This is particularly true if features are present in blazar spectra (breakdown of assumption a).  However, there is no evidence in EGRET data for such features.  In the cases where the other two assumptions (b and c) do not hold, as long as the dependencies are small, the decoupling of the shape and magnitude will still be approximately correct.  In VP07, possible correlations between spectral index and redshift and between spectral index and luminosity for blazars were investigated and no evidence for such correlations was found.  Nevertheless, absence of evidence is not equivalent to evidence of absence of a correlation,  and it cannot, as yet, be proven that the spectral index is independent of redshift and luminosity.  Since the gamma-ray emission from blazars is likely due to Inverse Compton emission which is related to emission at lower frequencies, a relationship between the gamma-ray spectral index and the gamma-ray luminosity is plausible and can be motivated,  for example, in the context of the blazar sequence (Fossati et al. 1998; Ghisellini et al. 1998).  On the other hand, the blazar sequence has been subsequently called into question with data from deeper blazar surveys (Giommi et al. 2005; Padovani 2007). Thus, it is difficult to interpret the VP07 result, especially based on current data alone; a further investigation of the blazar sequence  (as suggested in Maraschi et al. 2007) as well as posible correlations of luminosity with spectral index in the GeV energy range with GLAST data will offer further insight into the issue. However, we point out that systematic uncertainties in the spectral shape of the blazar contribution entering through correlations between blazar spectral index and blazar luminosity/redshift {\em weaker} than the VP07 constraints would be dominated by the uncertainties entering through our limited knwoledge of the ISID, or the systematic uncertainties in the observational determination of the EGRB. 
 
For now, we can instill more confidence in the reasoning behind the above arguments by applying our formalism to an independent SID for which a shape was determined using the full redshift-dependent, luminosity-dependent equation.
We applied Eq. \ref{tria} to the Stecker \& Salamon (1996a) SIDs\footnote{Note that appropriate adjustments had to be made: (1) separation between flaring and quiescent blazars as they did, and (2) rejection of unphysical spectral indices arising from a wide SID.} and found the shapes for flaring and quiescent blazars reported by the authors (although, of course, our formalism cannot reproduce the {\em amplitude} of the emission they reported).

\section{Inputs}

Traditionally, in order to evaluate the blazar contribution to the
EGRB, one would derive a gamma-ray luminosity function and a redshift
distribution and calculate the overall \emph{magnitude} of the
contribution.  However, in this analysis, we seek information about
the \emph{shape} of the blazar contribution rather than the overall
magnitude.  Thus, our only inputs are the ISIDs for different EGRET samples and simulated GLAST samples presented in VP07.  Since we do not include information about the magnitude of the blazar contribution to the EGRB, we have normalized our curves so that they always pass through the $83 {\rm \,
  MeV}$ best-guess measurement of the EGRB from EGRET data
($\left.E^2I_{\rm EGRB}\right|_{83 {\rm \, MeV}} = 1.5 \times 10^{-6}
{\rm \, GeV \, cm^{-2} s^{-1} sr^{-1}}$), as analyzed by Strong et al
(2004), in order to aid the visual comparison with the observed 
EGRB shape. This normalization is completely arbitrary as we could have instead chosen any one of the Strong et al. 2004 data points; we have chosen $83 \mbox{ MeV}$ since at this energy the observational
  statistical and systematic uncertainties are relatively low.


\section{Results}

\begin{figure}
\plotone{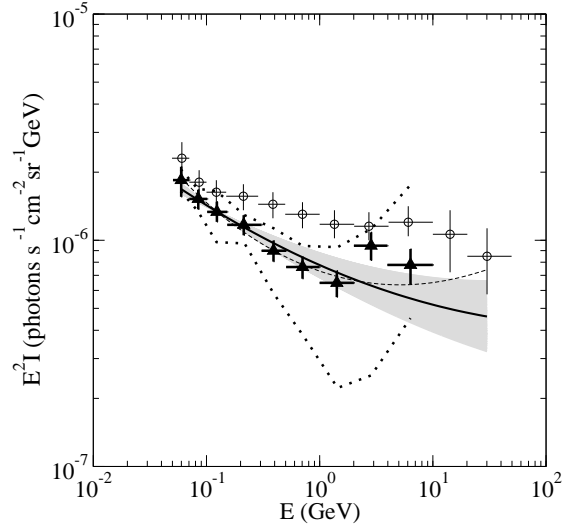}
  \caption{\label{fig1}
Spectral shape of the unresolved blazar emission. All curves have been normalized (arbitrarily) so that they pass through the 83 MeV point of the Strong et al.\ (2004) EGRB. Solid line: best-guess
spectrum based on a maximum-likelihood ISID determined using the Mattox et al.\ 2001 confident blazar sample. Grey region: spectral
shapes allowed for ISID parameters within the $1\sigma$ likelihood
contour. Filled triangles: Strong et al.\ (2004) EGRB
determination. Open circles: Sreekumar et al.\ (1998) EGRB
determination. Error bars are statistical errors only. Thick dotted
lines: Strong et al.\ (2004) EGRB systematics.  Thin dotted line:  SID determined in Stecker \& Salamon (1996a). }
\end{figure}

The spectral shape of the unresolved blazar emission for the ISID of
the Mattox et al.\ (2001) confident blazars sample (46 sources) is
plotted in Fig.\ \ref{fig1}.  The solid line is the best-guess
spectrum based on the maximum-likelihood Gaussian ISID for this sample
(mean $\alpha_0 = 2.27$ and spread $\sigma_0 = 0.2$) from VP07.  The
gray region represents spectral shapes derived from $(\alpha_0,
\sigma_0)$ pairs within the 1$\sigma$ contour of the ISID parameter
likelihood and illustrates our $1\sigma$ uncertainty in the unresolved
blazar spectral shape.  For comparison, we also plot datapoints with
(statistical) errors for the Sreekumar et al.\ (1998) (open circles)
and the Strong et al.\ (2004) (filled triangles)\footnote{The
  Sreekumar et al.\ (1998) and the Strong et al.\ (2004) EGRB
  determinations differ in the model used to subtract the diffuse
  emission of the Milky Way with the difference resulting from the fact that the Strong et al.\ (2004) determination
  is based on a Milky Way model which accounts for the GeV excess.}
determinations of the EGRB from EGRET data. We stress again that our
results are intended to be compared to the EGRET data only in shape
(relative intensity between different energy bins) and not in
amplitude, since the normalization of our curves does not carry any
information. 
 The thick dotted lines
indicate the Strong et al.\ (2004) systematics, entering through their
model of the Galaxy.  
The best-guess spectrum is only mildly curved in comparison to the
best-guess spectrum derived from the observed,
measurement-error--contaminated SID (as parametrized by Stecker \&
Salamon 1996a for the Mattox et al.\ 2001 confident blazar dataset;
shown as the thin dashed line in Fig.\ \ref{fig1}) in lieu of the
maximum-likelihood ISID.   

\begin{figure*}
\begin{center}
\resizebox{4.5in}{!}
{\plotone{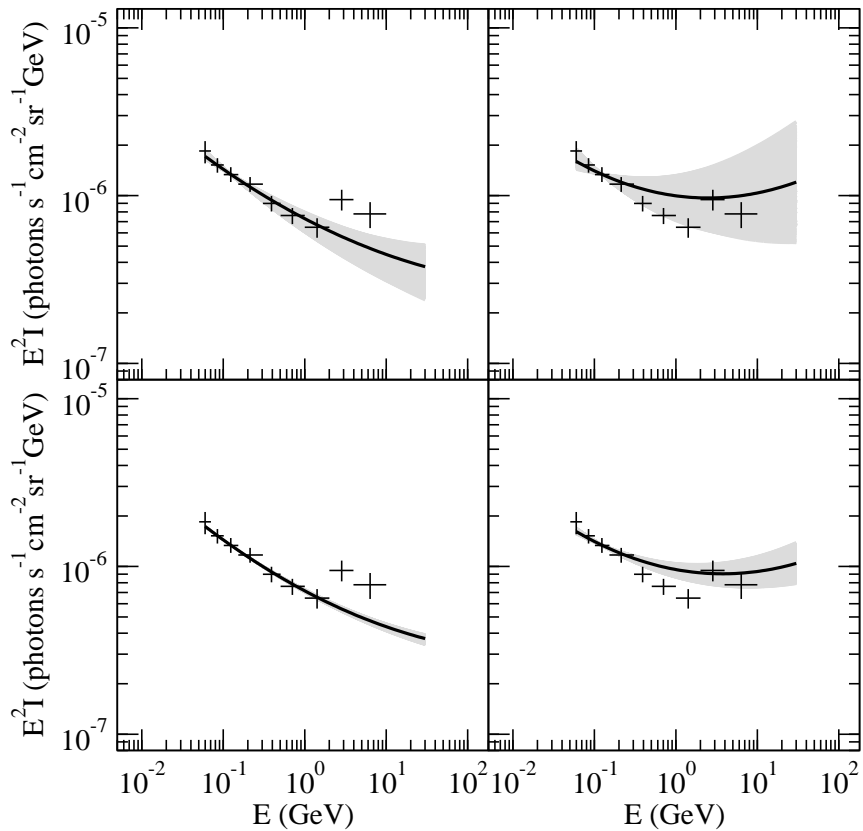}}
\caption{\label{fig2}
Predicted spectral shapes for FSRQs (left column) and BL Lacs (right column).  All curves have been normalized (arbitrarily) to the 83 MeV point of the Strong et al.\ EGRB. Solid lines represent the best-guess spectra and the gray regions, the $1\sigma$ spectral shape uncertainties. Data points:  Strong et al. (2004) EGRB determination (error bars represent statistical errors). Upper panel: EGRET data. Lower
panel: reduction of shape uncertainties in the GLAST era based on
the Dermer (2007) predictions for the numbers of detectable BL Lacs
and FSRQs. }
\end{center}
\end{figure*}

A most important result is that the 
constraints on the theoretical cumulative blazar spectrum are not
very strong. In fact, the theoretical uncertainties in the spectrum
entering through our limited understanding of the ISID are comparable
with the statistical uncertainties (errors on individual data points)
in the observed EGRB.  Worse yet, the large systematic and statistical
uncertainties in the determination of the EGRB impede any comparison
between the theoretical cumulative blazar spectrum and the EGRB. 
For this reason, no strong conclusions regarding the spectral
(in)consistency of the blazar collective spectrum with the observed
EGRB should be drawn based solely on EGRET data. Note that improvements in observations will \emph{not} automatically alleviate this concern.  Even if the systematic and statistical uncertainties in the data improve, if the uncertainties in the theoretical spectral shape remain as they are, strong conclusions about the blazar contribution to the EGRB will remain unattainable. However, if we were to
ignore systematics and take at face-value the upturn of the EGRET EGRB
at high energies indicated by Strong et al.\ (2004), our results
suggest that it is unlikely that EGRET blazars, as a population,
comprise the dominant contribution to the background at the highest
energies even if they do dominate at low energies. 

If BL Lacs and FSRQs have distinct intrinsic spectral index
distributions as indicated tentatively by EGRET data (Pohl et al.\
1997; Mukherjee et al.\ 1997; VP07), their cumulative unresolved
spectra will also differ.  The top row of Fig.\ \ref{fig2} shows the
spectral shapes of unresolved emission from FSRQs (left panel) and BL
Lacs (right panel) calculated from the respective VP07 ISIDs derived
from EGRET data. The solid lines again represent best-guess spectra
and the gray regions are $1\sigma$ uncertainties in the shape. The
observed EGRB shape (Strong et al.\ 2004 with statistical error bars)
is indicated with the crosses.

The best-guess spectra of the two populations have different shapes
with the cumulative emission from BL Lacs being generally harder than
that of FSRQs and having a more convex spectrum. However, due to the
poor number statistics of BL Lacs, their ISID is not well
constrained and this uncertainty is carried over to the emission
spectrum: the theoretical uncertainties in the unresolved BL Lac
spectral shape are much larger than the statistical uncertainties in
the data and comparable with the systematic uncertainties in the
observed EGRB.  Thus, no confident statement can be made regarding the
spectral (in)consistency of BL Lacs with the observed EGRB. On the
other hand, as there are many more FSRQs than BL Lacs detected by
EGRET, when all blazars are treated as a single population, both the
ISID and the unresolved emission spectrum closely resemble that of
FSRQs.  

If a similar ratio of FSRQs to BL Lacs is also present in the
unresolved blazar population (e.g.,  Dermer 2007) as in the resolved
EGRET blazar population, the shape of the blazar EGRB component will
mostly resemble that of Fig.\ \ref{fig1}. If on the other hand the
BL Lac fraction is much higher in unresolved blazars (e.g. Pohl et
al.\ 1997), then there may be an appreciable BL Lac contribution to
the EGRB, accounting, at least in part, for the upturn of the observed
EGRB tentatively suggested by EGRET data. GLAST observations will
greatly help in addressing this question as it is expected to resolve
between 1000 and 10,000 blazars, thus placing much stronger
constraints on the luminosity function and evolution of FSRQs and BL
Lacs.

GLAST observations will also allow a much more confident determination
of the FSRQ and BL Lac ISIDs resulting in corresponding improvements
in the determinations of the unresolved emission spectral shapes for these
populations. The lower row in Fig.\ \ref{fig2} shows the improvements
of our theoretical predictions for the FSRQ (left panel) and BL Lac
(right panel) unresolved spectra using ISIDs from the simulated GLAST
datasets of VP07. Note that this is simply a prediction of
  how much the uncertainties in the  
determinations of the spectral shapes will be reduced
  with increased number statistics and is not an actual prediction of
  the GLAST EGRB. Shape uncertainties in the GLAST era are reduced by a
factor of $\sim 3$. FSRQs and BL Lacs were assumed to follow the Dermer
(2007) luminosity functions which represent the most conservative
(lowest) predictions for the number of blazars that will be resolved
by GLAST. Note that the Dermer luminosity functions were
  \emph{not} used to determine the spectral shapes but solely to predict the numbers of BL Lacs and FSRQs GLAST will see, and thus, allowing us to estimate the reduction in the uncertainties on the
  ISIDs. If more 
blazars are, in fact, detected by GLAST, the ISIDs will be determined
with even greater confidence due to improved number statistics. Even
in the most conservative case, it is clear that GLAST observations
will place very tight constraints on the expected spectral shapes of
the unresolved emission of gamma-ray emitters - at least in the case
of blazar classes.  Additionally, GLAST should also be able to further
constrain the EGRB itself.  Thus, in light of GLAST, the spectral
shapes of the contributions of blazar populations would provide vital
information about whether those populations could, in principle,
explain all of the EGRB, or whether contributions from other gamma-ray
emitters are required.
 
\section{Discussion}\label{Disc}

We have calculated the expected spectral shape of the unresolved
gamma-ray emission from blazars under the assumptions that the ISIDs
of blazars do not evolve with redshift and are independent of blazar
luminosity and flaring state. We have also explicitly calculated the
$1\sigma$ uncertainty in the spectral shape entering through the
limited constraints on the blazar ISIDs derived from EGRET
data. Finally, we have predicted by how much these uncertainties will
be reduced if GLAST observations are used to determine the blazar
ISIDs. 

The unresolved emission spectral shape can be used as an indicator of
the potential importance of a given population's contribution to the
EGRB, and  it constrains the maximal contribution at high energies
relative to that at low energies.  If the curvature tentatively seen
in the observed EGRB is real, then a population with little such
curvature in its unresolved spectrum (such as the FSRQs) will not be
the dominant contributor to the EGRB at high energies even if it is
dominant at low energies. The unresolved BL Lac spectrum does seem to
be more convex than that of the FSRQs, but the level of uncertainty in the
spectral shape is high in this case because only a few BL Lacs were
detected by EGRET. However, GLAST observations will dramatically
improve our understanding of the blazar ISIDs and the associated
unresolved spectral shapes allowing us to use spectral shape
information to calculate the minimal additional contribution required
from other classes of sources to explain the observed EGRB spectrum.  

It should be stressed that here we have only calculated {\em
  unabsorbed} spectra. At energies higher than 10 ${\rm \, GeV}$,
gamma-ray absorption through interactions (pair production) with the
extragalactic background light becomes important (e.g., Salamon \&
Stecker 1998), and the spectral shape of any contribution to the EGRB
will be accordingly changed. We will return to this effect in a future
publication.  

\acknowledgements{We gratefully acknowledge enlightening discussions
with Brian Fields, Julie McEnery, Angela Olinto, Martin Pohl, Jenny
Siegal-Gaskins, Floyd Stecker, and Kostas Tassis. This work was supported in part by the Kavli Institute for Cosmological Physics at the University of Chicago through grants NSF PHY-0114422 and NSF PHY-0551142 and an endowment from the Kavli Foundation and its founder Fred Kavli. T.M.V. was supported by an NSF Graduate
Research Fellowship.


\begin{thebibliography}{}

\bibitem[Brecher \& Burbidge(1972)]{1972ApJ...174..253B} Brecher, K., \& 
Burbidge, G.~R.\ 1972, \apj, 174, 253 

\bibitem[de Boer et al.(2004)]{db05} de Boer, W., Sander, C., Gladyshev, A.~V.\ \& Kazakov, D.~I. 2005, A\&A 444, 51  
\bibitem[Dermer(2006)]{D06} Dermer, C.~D.\ 2007, \apj, 659, 958

\bibitem[Fossati et al.(1998)]{F98} Fossati, G., Maraschi, L., Celotti, A., Comastri, A., \& Ghisellini, G.\ 1998, \mnras, 299, 433

\bibitem[Ghisellini et al.(1998)]{G98} Ghisellini, G., Celotti, A., Fossati, G., Maraschi, L., \& Comastri, A.\ 1998, \mnras, 301, 451

\bibitem[Hartman et al.(1999)]{har99} Hartman, R.~C., et al.\ 
1999, \apjs, 123, 79 

\bibitem[Lichti et al.(1978)]{lic78} Lichti, G.~G., Bignami, G.~F., \& Paul, J.~A.\ 1978, ApSS, 56, 403

\bibitem[Maraschi et al.(2007)]{mar07} Maraschi, L., Ghisellini, G., \& Tavecchio, F.\ 2007, AIP Conference Proceedings, 921, 160

\bibitem[M{\"u}cke \& Pohl(2000)]{mp00} M{\"u}cke, A., \& 
Pohl, M.\ 2000, \mnras, 312, 177 

\bibitem[Mukherjee et al.(1997)]{Metal97} Mukherjee, R., et al. \ 1997,
\apj, 490, 116

\bibitem[Mukherjee \& Chiang(1999)]{mc99} Mukherjee, R., \& 
Chiang, J.\ 1999, Astroparticle Physics, 11, 213 

\bibitem[Narumoto \& Totani(2006)]{NT06} Narumoto, T., \& 
Totani, T.\ 2006, \apj, 643, 81 

\bibitem[Padovani et al.(1993)]{pad93} Padovani, P., 
Ghisellini, G., Fabian, A.~C., \& Celotti, A.\ 1993, \mnras, 260, L21 

\bibitem[Padovani(2007)]{pad07} Padovani, P.\ 2007, \apss, 309, 63


\bibitem[Pavlidou \& Fields(2002)]{pf02} Pavlidou, V., \& 
Fields, B.~D.\ 2002, \apjl, 575, L5 

\bibitem[Pavlidou et al.(2007)]{petal07} Pavlidou, V., Siegal-Gaskins, J.~M, Fields, B.~D., Olinto, A.~V., \& Brown, C.\ 2007, {\it sumbitted to ApJ}
\bibitem[Pohl et al.(1997)]{p97} Pohl, M., Hartman, R.~C., 
Jones, B.~B., \& Sreekumar, P.\ 1997, \aap, 326, 51 

\bibitem[Salamon \& Stecker(1998)]{sal98} Salamon, M.~H., \& 
Stecker, F.~W.\ 1998, \apj, 493, 547 

\bibitem[Shen et al. (2006)]{she06} Shen, R., Kumar, P., \& Robinson, E.~L.\ 2006, MNRAS, 371, 1441.

\bibitem[Stecker \& Salamon(1996)]{ste96a} Stecker, F.~W., \& 
Salamon, M.~H.\ 1996a, \apj, 464, 600 
 
\bibitem[Stecker \& Salamon(1996)]{ste96b} Stecker, F.~W. \& 
Salamon, M.~H.\ 1996b, Physical Review Letters, 76, 3878 

\bibitem[Strong et al.(2004)]{smr04} Strong, A.~W., 
Moskalenko, I.~V., \& Reimer, O.\ 2004, \apj, 613, 956 

\bibitem[Thompson et al.(1995)]{tetal95} Thompson, D.~J.\ et al.\ 1995, ApJS, 101, 259

\bibitem[Venters \& Pavlidou (2007)]{vp07} Venters, T.~M. \& Pavlidou, V., ApJ {\it in press}, arXiv:0704.2417

\end{thebibliography}
\end{document}